\documentclass[aps,prl,superscriptaddress,twocolumn,showpacs]{revtex4}

\renewcommand*{\[}{\begin{equation}}

\renewcommand*{\]}{\end{equation}}

\usepackage{graphicx}

\usepackage{hyperref}

\begin{document}

\title{Momentum mapping of continuum electron wave packet interference}


\author{Weifeng Yang}

\affiliation{Department of Physics, College of Science, Shantou University,
Shantou, Guangdong 515063, China}

\author {Huatang Zhang}

\affiliation{Department of Physics, College of Science, Shantou University,
Shantou, Guangdong 515063, China}

\author{Cheng Lin}

\affiliation{Department of Physics, College of Science, Shantou
University, Shantou, Guangdong 515063, China}

\author{Jingwen Xu}

\affiliation{Department of Physics, College of Science, Shantou
University, Shantou, Guangdong 515063, China}

\author{Zhihao Sheng}

\affiliation{Department of Physics, College of Science, Shantou
University, Shantou, Guangdong 515063, China}

\author{Xiaohong Song}
\email{songxh@stu.edu.cn}

\affiliation{Department of Physics, College of Science, Shantou University,
Shantou, Guangdong 515063, China}

\author{Shilin Hu}
\affiliation{CAPT, HEDPS, and IFSA
Collaborative Innovation Center of MoE College of Engineering,
Peking University, Beijing 100084, China}

\affiliation{Institute of Applied Physics and Computational
Mathematics, P. O. Box 8009, Beijing 100088,China}

\author{Jing Chen}
\email{chen\underline{ }jing@iapcm.ac.cn} \affiliation{CAPT, HEDPS, and IFSA
Collaborative Innovation Center of MoE College of Engineering,
Peking University, Beijing 100084, China}

\affiliation{Institute of Applied Physics and Computational Mathematics, P. O. Box 8009, Beijing 100088,China}

\date{\today}

\begin{abstract}

We analyze the two-dimensional photoelectrons momentum
distribution of Ar atom ionized by midinfrared laser pulses and
mainly concentrate on the energy range below $2U_{p}$. By using a
generalized quantum trajectory Monte Carlo (GQTMC) simulation and
comparing with the numerical solution of time-dependent
Schr\"{o}dinger equation (TDSE), we show that in the deep
tunneling regime, the rescattered electron trajectories plays
unimportant role and the interplay between the
intra-cycle and inter-cycle results in a ring-like interference
pattern. The ring-like interference pattern will mask
the holographic interference structure in the low longitudinal
momentum region. When the nonadiabatic tunneling contributes
significantly to ionization, i.e., the Keldysh parameter
$\gamma\sim1$, the contribution of the rescattered electron
trajectories become large, thus holographic interference pattern
can be clearly observed. Our results help paving the way for
gaining physical insight into ultrafast electron dynamic process
with attosecond temporal resolution.

\end{abstract}


\maketitle

\section{I. Introduction}

Atomic photoionization under intense laser irradiation is a
fundamental process in strong-field light-matter interaction. The
physical picture is understood by the simple man's model
\cite{Corkum3}. Within this model, the electron is released first
from its parent atomic core by ionization, then is accelerated in
the laser field and redirected to the parent ion, and finally
re-collides with the parent ion. Usually, the ionization process
is divided into two regimes: multiphoton ionization and tunneling
ionization. The Keldysh parameter,
$\gamma=\sqrt{\frac{I_{p}}{2U_{p}}}$ ($I_{p}$ is the ionization
potential, and $U_{p}=\frac{I}{4\omega^{2}}$ denotes the
ponderomotive energy, where $\emph{I}$ is the laser intensity and
$\omega$ is the angular frequency), is an indicator as to
distinguishing these two ionization regimes \cite{Keldysh1}. When
$\gamma$ is much less than $1$, the ionization process is in the
tunneling regime, where quasi-static approximation is valid. When
$\gamma$ is much larger than $1$, it is in the multiphoton
ionization regime. When $\gamma \sim 1$, it is considered to be a
transition from the tunneling regime to the multiphoton regime,
and the barrier in the combined Coulomb and laser field potential
changes significantly during tunneling \cite{Yu33}.

Interference is a very important concept of coherent matter waves
which has been extended successfully to explore highly nonlinear
quantum-mechanical phenomena in strong-field light-matter
interaction. The interference of the two electron wave packets
(EWPs) ionized with exactly one optical cycle relative delay
reaching the same final momentum gives rise to above-threshold
ionization (ATI) rings, i.e., the intercycle interference, that
are spaced by the energy of one photon in photoelectron spectrum.
In addition to the intercycle interference, a temporal double-slit
pattern can be verified as signature of EWP interference emitted
from the successive maxima of the absolute value of the electron
field, which is the intracycle interference
\cite{Lindner2005,Gopal2009}. The interference pattern of the
interplay between intra- and intercycle interferences in
multicycle photoelectron spectra has been identified as
diffraction pattern for a time grating
\cite{Gopal2009,DiegoPRAR2010,Diego2010, Xie2012, Remetter2006}.
Recently, a holographic structure has been observed in
photoionization and it is demonstrated to be the interference
between the direct and rescattered EWPs ionized within the same
quarter-cycle of the laser pulse
\cite{Huismans2011,Marchenko2011,Hickstein2012,Yang21,Meckel2014,Song2016}.
In the total photoelectron momentum distribution spectrum, all
these interference processes will interplay with each other, and
various interference patterns will mix together. As a result,
establishing an unambiguous one-to-one relationship between
certain interference pattern and the corresponding electronic
dynamic process is essential for retrieving the information of
electronic dynamics from the measured photoelectron momentum
spectrum.

On the other hand, with the development of intense mid-IR sources,
experimental probing deep into the tunneling regime has become
possible. Using a high repetition rate OPCPA, Keldysh parameters
approaching $\gamma\sim0.1$ can be achieved \cite{Pullen2014}. In
this regime, unexpected low energy structure, very low energy
structure, and even zero energy structure have been observed
\cite{Blaga2009NatPhys, Quan2009PRL,Wu2012PRL,Dura2013}. All these
experimental results and the following theoretical analysis
greatly advances people's understanding in this field. In the
original paper of strong-field photoelectron holography, tunneling
ionization had been assumed to be essential for the holographic
interference (in that experiment, $\gamma=0.76$)
\cite{Huismans2011}. Subsequently, further investigations indicated
that the holographic interference pattern can also be observed
under the conditions belong to the multiphoton regime ($\gamma>1$)
\cite{Marchenko2011}. However, whether the photoelectron hologram
can be observed in deep tunneling regime (i.e., $\gamma\ll1$) has
not yet been analyzed.

In the present work, we analyze the photoelectron angular
distributions (PADs) in atomic ATI with midinfrared laser pulses.
A profound ring-like interference pattern is identified by both of
time-dependent Schr\"{o}dinger equation (TDSE) and generalized
quantum-trajectory Monte Carlo (GQTMC) simulations in the deep
tunneling regime. Within the description of the GQTMC, the
ring-like interference pattern is demonstrated to be the
superposition between the intra- and inter-cycle interference. The
center of the ring-like interference pattern lies in where
separations between adjacent temporal double-slit interference
fringes are nearly the same as those of the ATI rings. Moreover,
Coulomb potential plays a negligible role on the formation of the
ring-like interference pattern. The existent of the ring-like
interference pattern will mask the holographic interference
pattern in the low final longitudinal momentum range, so that the
holographic interference pattern can only be observed in the high
final longitudinal momentum range. As a result, we identify that
deep tunneling is not a appropriate condition for observing the
holographic interference pattern. In the nonadiabatic regime, the
contribution of the rescattering electron trajectories will
increase, so the holographic interference will be clearly
observed.

This paper is organized as follows. In Sec. II we introduce the
theoretical methods including the numerical solutions of the TDSE
and the GQTMC model. In Sec. III, firstly, we show different
characteristics of the interference structures in PADs in different laser parameter regions
using the TDSE and GQTMC simulations. Secondly, the underlying
mechanism of the ring-like pattern is discussed based on the GQTMC
method. Moreover, the intra-, intercycle interference and the
Coulomb potential effects on the interference pattern are
discussed. We summarize our results and conclude in Sec. IV.

\section{II. Theoretical models}

In this section, we summarize the numerical solution of the TDSE
and the GQTMC method. The numerical solution of the TDSE is
considered to be exact and can be used as a benchmark for
assessing the validity of the GQTMC method.

\subsection{The numerical solution of the time-dependent
Schr\"{o}dinger equation}

We consider an atom interacting with a linearly polarized laser
field within the single active electron approximation. The
electric field of the laser pulse is
\begin{equation}
\textbf{E}(t)=E_{0}f(t)\textmd{cos}(wt)\hat{z},
\end{equation}
where $\hat{z}$ is the laser polarization direction, with $f(t)$
the pulse envelope function. $E_0$ is peak field strength. We
solve the atomic TDSE

 \begin{eqnarray}
i\frac{\partial}{\partial t}\Psi(\textbf{r},t)=\{\frac{\textbf{p}^{2}}{2}+\textbf{p}\cdot\textbf{ A(t)}+\mathbf{V}(\mathbf{r})\}\Psi(\textbf{r},t).
\end{eqnarray}
Here, $\textbf{p}$ is the momentum, $\textbf{V}(\textbf{r})$ is
the atomic potential of Ar,
$\textbf{A(t)}=-\int_{0}^{t}\textbf{E}(t')dt'$ is the vector
potential, and \textbf{r} the position of the electron. The exact
time evolution of the wave function $\Psi(t)$ is evaluated by
using the split-operation method in energy representation
\cite{Tong1997,Yang2012}. The space is split into two parts, i.e., the
inner and outer region where the atomic potential becomes
negligible compared to the kinetic energy. When the time-dependent
wave function in space reaches the outer region, we project the
outer region wave function on Volkov states to obtain the momentum
distribution \cite{Yang21,Tong2006}.

\subsection{The generalized quantum trajectory Monte Carlo
method}

To explore the physical reason of the TDSE results, we apply a
GQTMC method \cite{Song2016} based on the nonadiabatic ionization theory
\cite{Yudin2001,Perelomov1966}, classical dynamics with combined
laser and Coulomb fields \cite{Brabec1996,Hu1997,Chen2000}, and
the Feynman's path integral approach \cite{Salieres2001,Li2014}.
The ionization rate is given as:

\begin{equation}
\Gamma(t)=N(t)\exp(-\frac{E_{0}^{{2}}f^{{2}}(t)}{\omega^{{3}}}\Phi(\gamma(t),\theta(t))),
\end{equation}
where $\theta(t)$ is the phase of the laser electric field. For
convenience of analysis, the laser pulse envelope $f(t)$ is
half-trapezoidal, constant for the first four cycles and ramped
off linearly within the last two cycles. The preexponential factor
is
\begin{eqnarray}
&N(t)=A_{n^{*},l^{*}}B_{l,|m|}(\frac{3\kappa}{\gamma^{{3}}})^{\frac{{1}}{{2}}}CI_{p}(\frac{2(2I_{p})^{{3/2}}}{E(t)})^{2n^{*}-|m|-1} \nonumber \\
\label{eq:12}\\
&\kappa=\ln(\gamma+\sqrt{\gamma^{{2}}+1})-\frac{\gamma}{\sqrt{\gamma^{{2}}+1}}\nonumber
\end{eqnarray}
where the coefficient $A_{n^{*},l^{*}}$ and $B_{l,|m|}$ coming from the radial and angular part of the wave function, are given
by Eq.(2) of Ref. \cite{Yudin2001}.
$C=(1+\gamma^{{2}})^{|m|/2+3/4}A_{m}(\omega,\gamma)$ is the Perelomov-Popov-Terent'ev correction to the quasistatic
limit $\gamma\ll1$ of the Coulomb preexponential factor with $A_{m}$ given by Eqs. (55) and (56) of Ref. \cite{Perelomov1966}. The
tunnelled electrons have a Gaussian distribution on the initial
transverse momentum
$\Omega(v_{r}^j,t_{0})\propto[v_{r}^j\sqrt{2I_{p}}/|E(t_{0})|]\exp[\sqrt{2I_{p}}(v_{r}^j)^{2}/|E(t_{0})|]$.

The coordinate of the tunnel exit shifts toward the atomic core
due to the nonadiabatic effects \cite{Perelomov1966}, and the
tunnel exit point is
\begin{equation}
{Z_0}=\frac{2I_p}{E(t_0)}(1+\sqrt{1+\gamma^{2}(t_0)})^{-1}
\end{equation}
Thereafter, the classical motion of the electrons in the combined
laser and Coulomb fields is governed by the Newtonian equation:
\begin{equation}
\frac{d^{2}}{dt^{2}}\mathbf{r}=-\mathbf{E}(t)-\bigtriangledown(\mathbf{V}(\mathbf{r})).
\end{equation}
According to the Feynman's path integral approach, the phase of
the $\emph{j}$th electron trajectory is given by the classical
action along the trajectory
\begin{equation}
{S_j} (\mathbf{p},t_0 )=\int_{t_0}^{+\infty}\{\mathbf{v_p^{2}}(\tau)/2+I_p-1/|\mathbf{r}(t)| \}d\tau
\end{equation}
where $\textbf{p}$ is the asymptotic momentum of the $\emph{j}$th
electron trajectory. The probability of each asymptotic momentum
is determined by
\begin{equation}
{|\Psi_\mathbf{p}|}^{2}=|\sum_j\sqrt{\Gamma(t_0,v_{r}^j) }\rm{exp}(\it{-iS_j})|^{\textmd{2}}.
\end{equation}

With GQTMC method, as showing in the following section, one can
reproduce the TDSE results and extract all the information about
the electron trajectory including the initial ionization phase and
velocity. Moreover, by GQTMC method, one can reconstruct momentum
distribution with the photoelectrons from special subcycle time
windows, which is in favour of exploring the interference and
Coulomb potential effects.

\section{III. Results and discussion}

Using the TDSE [Figs. 1 (a) and (c)] and GQTMC [Figs. 1 (b) and
(d)], we have calculated PADs of ionization from Ar atom in
linearly polarized laser fields with different laser parameters
(1850 nm, $1.05\times10^{14}$W/cm$^{2}$, corresponding to
$\gamma=0.48$, and 1300 nm, $0.4\times10^{14}$W/cm$^{2}$,
corresponding to $\gamma=1.11$). The GQTMC simulations agree well
with the TDSE results in both of the tunneling ionization regime
and the transition regime. Both of the TDSE and GQTMC simulations
show that, in the tunneling ionization regime, in addition to the
ATI rings which center at zero, another ring-like interference
pattern centers at about $p_{z}\sim0.56a.u.$ and $p_{r}=0$ can be
clearly observed. However, in the transition regime where $\gamma
\sim 1$, the ring-like interference pattern disappears, whereas
the holographic ``fork'' interference structure can be identified
in the PADs. Next we will discuss the physical mechanism of the
ring-like interference pattern and the conditions required for the
appearance of these different interference patterns.

\begin{figure*}[htbp]
\centering
\includegraphics[width=0.5\textwidth,height=0.45\textwidth,angle=0]{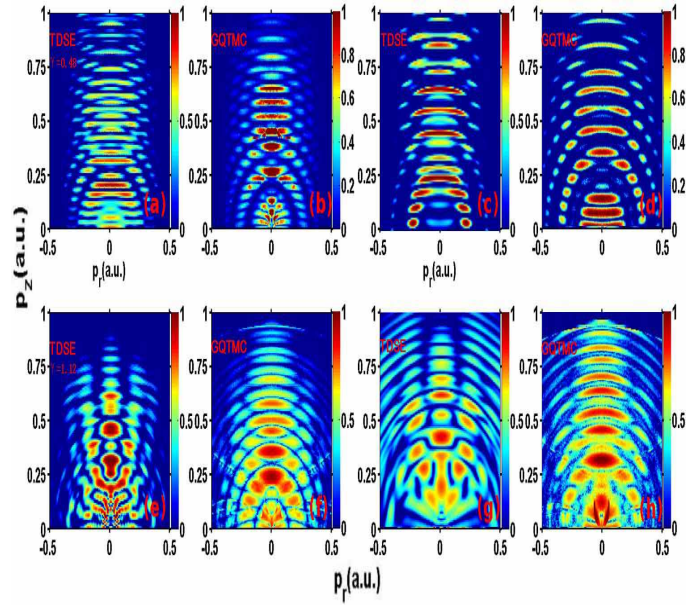}
\caption{(color online). Simulated two-dimensional
photoelectron momentum spectra of Ar atom. Left panel:
TDSE results; right panel: GQTMC simulations.
(a)-(c): ${\gamma}=0.55$, $I=7.0\times{10^{13}}$ W/cm$^{2}$, $\lambda=1700$
nm. (d)-(f): ${\gamma}=1.11$, $I=1.5\times{10^{13}}$ W/cm$^{2}$,
$\lambda=1700$ nm.}
\label{fig:false-color}
\end{figure*}

With the help of the GQTMC back analysis of the PADs, we can
disentangle different contributions of photoelectrons emitted from
different time windows and analyze the interplay among them. We
find the photoelectrons contributing to the ring-like fringes come
from at least three sub-cycle time windows, which are labeled as
A, B and C in Fig. 2(a). It is found that in the time windows A or
B, there are two kinds of typical trajectories: the rescattering
trajectory (trajectory R in Fig. 2(b)) and the so-called indirect
trajectory (trajectory ID in Fig. 2(c)). The main distinction
between these two kinds of trajectories lies in that the wave
packet of the trajectory ID does not interact with the parent ion
when it comes back close to $z=0$. The interference between these
two types of trajectories from single A or B will induce the
fork-like holographic interference pattern. In window C, there is
only one kind of typical trajectory: direct trajectory (trajectory
D in Fig. 2(d)). It has been well-known that the intercycle
interference of EWPs liberated with a relative time delay of one
optical cycle (for example, the interference between EWPs coming
from window A and B) will induce a series of ATI rings separated
by one photon energy. While the intracycle interference of EWPs
from window A and C induces a temporal double-slit interference
pattern which has been studied both experimentally and
theoretically \cite{Lindner2005}.

\begin{figure*}[htbp]
\centering
\includegraphics[width=0.6\textwidth,height=0.4\textwidth,angle=0]{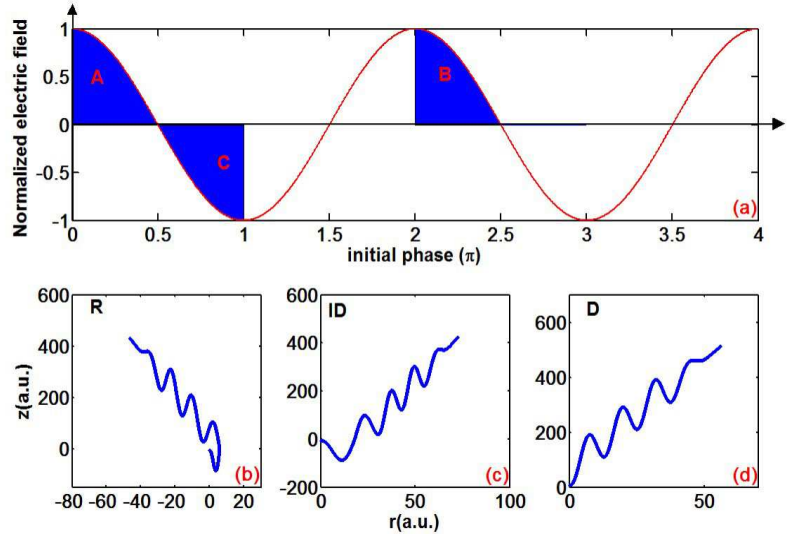}
\caption{(color online). Upper row (a): The electric field of laser. Lower row:(b)
the collections of typical trajectory R,(c) the collections of
typical trajectory ID,(d) the collections of typical trajectory
D.} \label{fig:false-color}
\end{figure*}

Fig. 3 shows reconstructed photoelectron momentum spectra from
different kinds of interference. Figs. 3(a) and (d) show the
temporal double-slit interference pattern reconstructed with EWPs
released from time windows A and C; Figs. 3(b) and 3(e) are the
ATI interference pattern reconstructed with EWPs from time windows
A and B; whereas Figs. 3(c) and (f) are photoelectron spectra
reconstructed with EWPs released from all these three time windows
A, B, and C. One can see that, compared with the outgoing ATI
rings, the temporal double-slit interference pattern is a
ring-like structure centered a much higher final momentum $p_{z}$.
Moreover, the separations of neighbor ring-like fringes are
unequal and, on the contrary to the ATI rings, gradually increase
with energy in the momentum distribution map.
\begin{figure*}[htbp]
\centering
\includegraphics[width=0.6\textwidth,height=0.35\textwidth,angle=0]{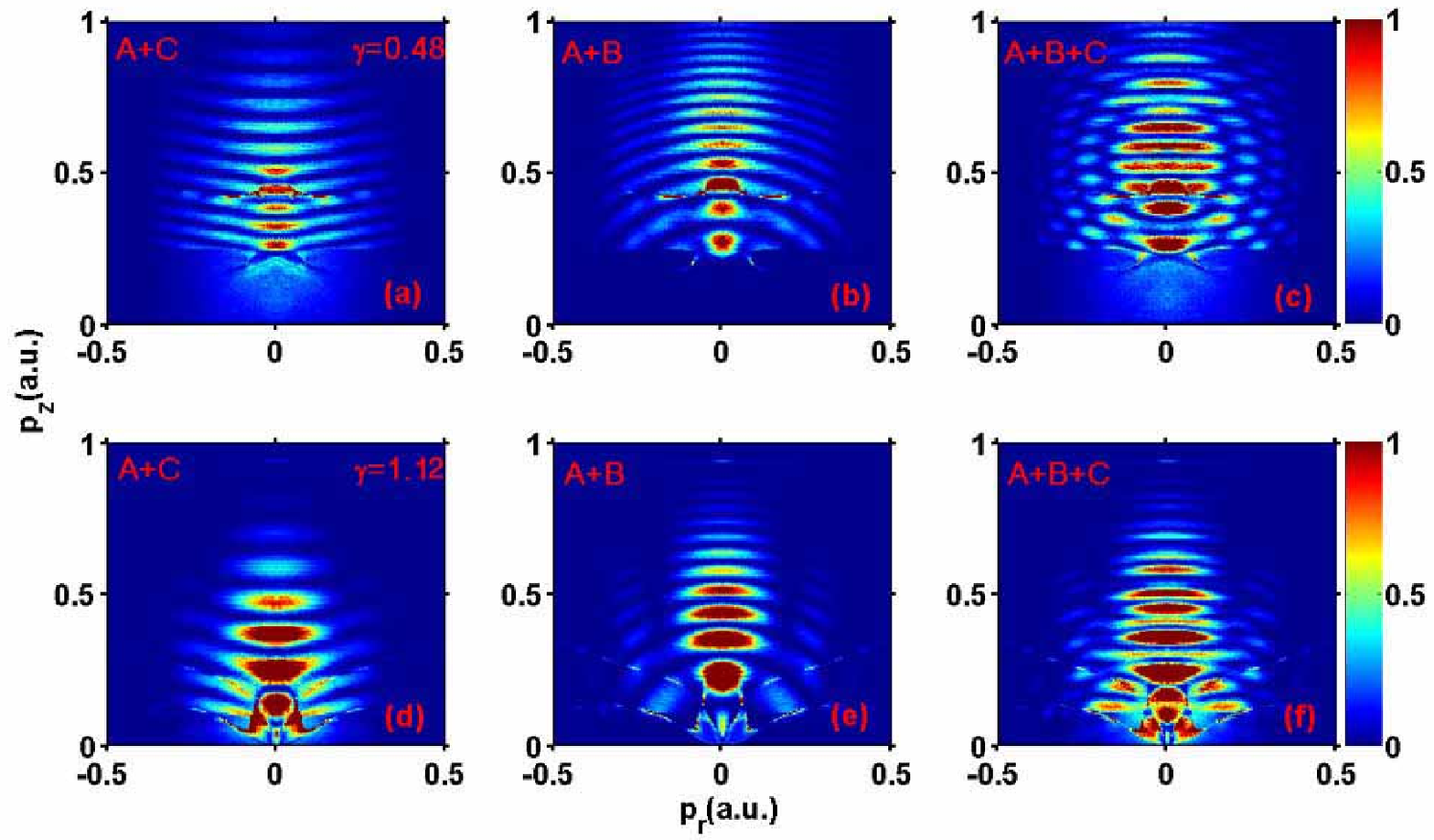}
\caption{(color online). Reconstructed photoelectron momentum spectra with EWPs
from different time windows: (a) and (d) with EWPs from time
windows A and B; (b)and (e) with EWPs from time windows A and C;
(c) and (f) with EWPs from time windows A, B, and C.
(a)-(c):${\gamma}=0.48$. (d)-(f): ${\gamma}=1.11$.}
\label{fig:false-color}
\end{figure*}

To show more clearly the influence of the Coulomb potential, we
further present in Fig. 4 the results without considering the
Coulomb potential. Comparing with Fig. 3, we can see that the
impact of the Coulomb potential is reflected mainly in: (i) it
distorts substantially the PADs in the low momentum ranges; (ii)
it introduces the rescattered trajectory, i.e., the trajectory R.
Without the Coulomb potential, the trajectory R will not exist,
and so does the holographic interference. In both Figs. 3(a) and
3(b), the holographic interference structures can be hardly seen,
which means that the contribution of trajectory R is quite small
under the laser condition $\gamma=0.48$. Most interestingly, when
we reconstruct the momentum distribution of photoelectrons from
all the three windows, the ring-like interference pattern comes
out in both cases with and without considering the Coulomb
potential (see Fig. 3(c)and Fig. 4(c)], demonstrating that the
Coulomb potential and the so-induced trajectory R play an
negligible role in this interference structure.

\begin{figure*}[htbp]
\centering
\includegraphics[width=0.6\textwidth,height=0.35\textwidth,angle=0]{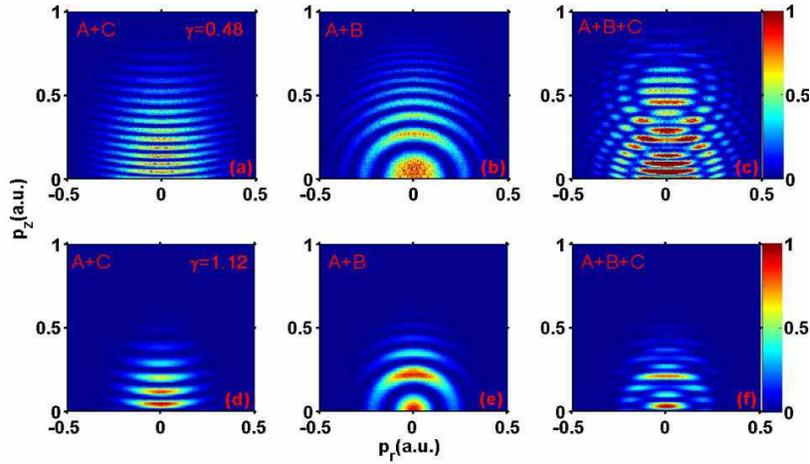}
\caption{(color online). Same with Fig.3 but without considering the Coulomb potential.}
\label{fig:false-color}
\end{figure*}

In the case of $\gamma=1.11$, the situation is quit different. The
contribution of trajectory R can already be discerned in the
intracycle temporal double-slit interference structure. In the
intercycle interference photoelectron momentum distribution (see
Fig. 3 (e)), both the fork-like holographic interference pattern
along with the ATI rings can be clearly distinguished. Remarkably
different from the case of $\gamma=0.48$, the momentum
distribution of photoelectrons from these three windows cannot
form ring-like structure. On the contrary, the holographic
interference structure can be clearly observed (see Fig. 3 (f)).
This demonstrates that the rescattered trajectory R indeed contributes
importantly to holographic interference in the total photoelectron momentum distribution in
this laser condition.

\begin{figure*}[htbp]
\centering
\includegraphics[width=0.65\textwidth,height=0.35\textwidth,angle=0]{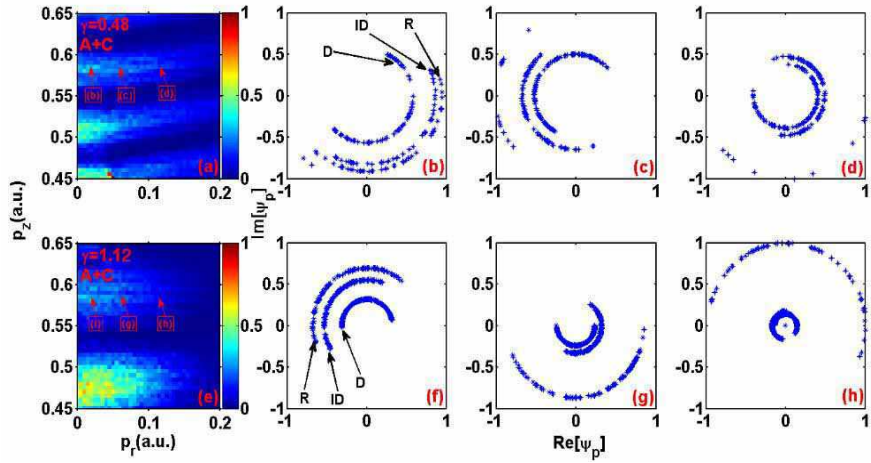}
\caption{(color online). (a) and (e) present zooms into Figs. 3(a) and (d) with 3
sampling points indicated. (b-d) and (f-h) show the probabilities
of the summed trajectories $\Psi_\mathbf{p}$ in the complex
plane.} \label{fig:false-color}
\end{figure*}

\begin{figure*}[htbp]
\centering
\includegraphics[width=0.65\textwidth,height=0.35\textwidth,angle=0]{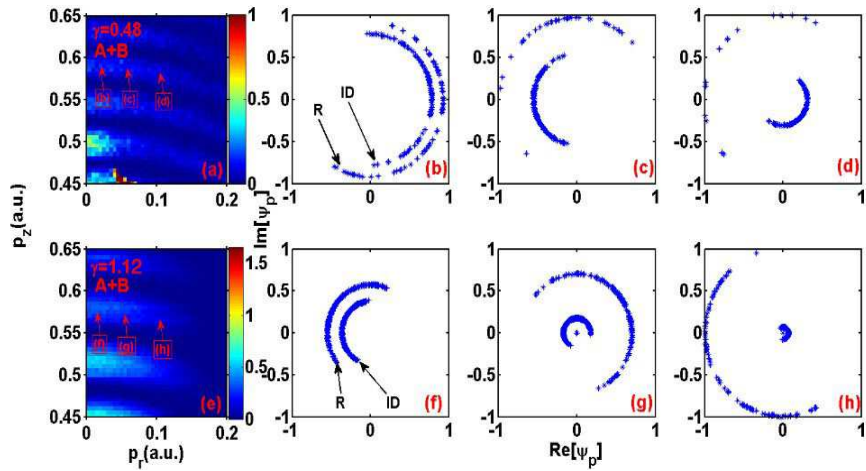}
\caption{(color online). Same with Fig. 5 but for three sampling points in Figs. 3(b) and (e).}
\label{fig:false-color}
\end{figure*}

The above analysis can be further demonstrated by the statistical
trajectory-based analysis in the complex plane \cite{TMYan2013}.
Figs. 5(a) and (e) present zooms into Figs. 3(a) and (d). Figs.
5(b-d) and (f-h) show the probabilities of the trajectories
$\Psi_\mathbf{p}$ in the complex plane (the radius represents the
weight $|\sqrt{\Gamma(t_0,v_{r}^j)}|$ and the angle represents the
phase $S_j$ of each trajectory) for 3 sampling points indicated in
Figs. 5(a) and 5(e), respectively. Fig. 6 shows the same analysis
but for the intracycle interference (Figs. 3 (b) and (e)). It can
be seen that for each sampling points in the intracycle
interference pattern, there are three arcs which correspond to the
three different kinds of electron trajectories (trajectories D, ID
and R), whereas for intercycle interference, only two arcs
corresponding trajectories D and R contribute to the interference
pattern. The outermost arc corresponds to the rescattered
trajectory R. When the two arcs align in the same direction,
constructive interference occurs, otherwise, the opposite
alignment leads to destructive interference. By comparing the two
cases with different laser conditions, we can see that, for
$\gamma=0.48$, the number of the rescattered trajectory R is
greatly reduced. On the contrary, for $\gamma=1.11$, the
rescattered trajectory R contributes greatly to the total momentum
spectrum. As a result, holographic interference structure can be
clearly distinguished from the the total momentum distribution
spectra (see Figs. 1 (c) and (d)). All these are consistent with
the above analysis. Therefore, we can conclude that the
rescattered trajectory R is very vital: when the contribution of
rescattered trajectory R increases, the holographic interference
pattern will be clearly visible, otherwise, the ring-like
interference structure will be formed due to the interplay between
the intracycle and intercycle interference of direct and indirect
electron trajectories.

\begin{figure*}[htbp]
\centering
\includegraphics[width=0.6\textwidth,height=0.4\textwidth,angle=0]{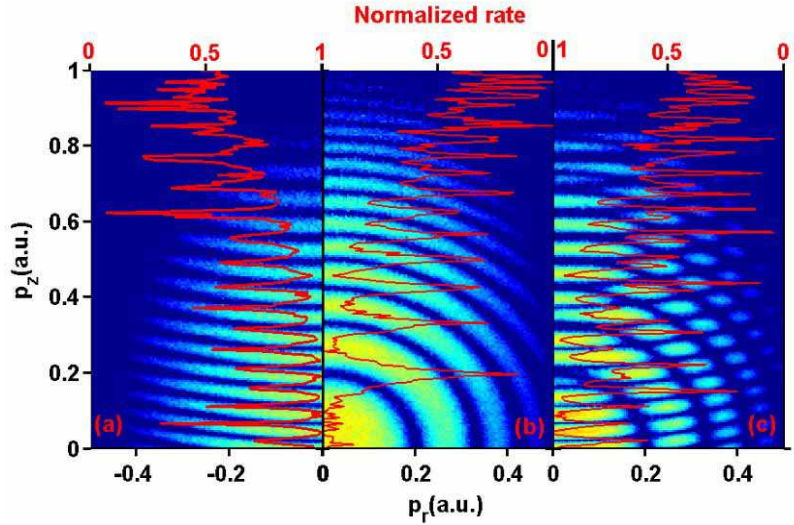}
\caption{(color online). (a)temporal double-slit interference patten, (b) ATI
rings, (c) the total momentum distribution spectrum reconstructed
with EWPs from three time windows A, B and C. The red lines
indicates the momentum spectrum for $p_{r}=0 a.u.$}
\label{fig:false-color}
\end{figure*}

\begin{figure*}[htbp]
\centering
\includegraphics[width=0.35\textwidth,height=0.6\textwidth,angle=0]{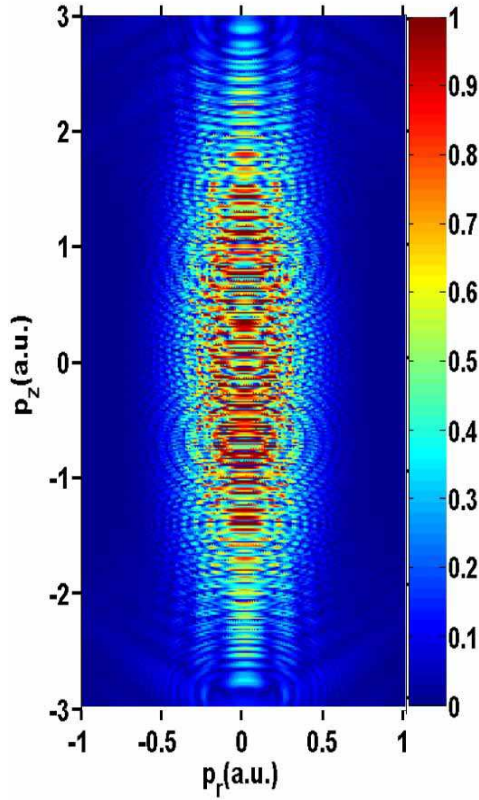}
\caption{(color online). TDSE simulation in deep tunneling regime: ${\gamma}=0.28$, $I=2.76\times{10^{14}}$ W/cm$^{2}$, $\lambda=2000$
nm.}
\label{fig:false-color}
\end{figure*}

Since the Coulomb potential plays an negligible role in the
ring-like interference, to shed more light on the physical
origination of the ring-like interference pattern, our further
analysis will be based on our GQTMC simulations without
considering the Coulomb potential. Figs. 7(a),(b) and (c) show
the temporal double-slit interference patterns, ATI rings,
and the total momentum distribution spectrum, respectively. The
red lines indicate the momentum spectrum for $p_{r}=0$. We find
that at small momentum range (i.e., $p_{z}<0.56 a.u.$), the
separations between adjacent temporal double-slit interference
fringes are smaller than those of the ATI rings, as a result, the
total photoelectron spectrum clearly shows the modulation of the
intracycle double-slit interference by the intercycle
interference. At around $p_{z}\sim0.56 a.u.$, the separations
between adjacent temporal double-slit interference fringes are
nearly the same as those of the ATI rings. In the total momentum
distribution spectrum, this forms the center of the ring-like
interference pattern. Above this momentum region, the separations
of temporal double-slit interference fringes are larger than those
of the ATI fringes, so the double-slit interference fringes are
imprinted a modulation envelope of the ATI fringes.
We collect the phases of indirect and direct electron trajectories from time windows A, B and C resulting in the ring-like pattern, and found that the phase difference between electron trajectories from A and B is approximately equal to that from A and C, which confirms the above analysis of Fig. 7.
In ref.\cite{DiegoPRAR2010}, it has been shown through a 1D ``simple
man's" model that the energy separation between adjacent peaks
will reach a maximum and then decrease with increasing
photoelectron energy. It means that with increasing $p_{z}$, there
will be more than one chance that the separations between adjacent
fringes for the two interference processes become nearly the same,
and so there would be more than one ring-like interference
structures in the total momentum distribution spectrum if the
scattered trajectory R do not contribute significant.

In Fig. 8, we further show the TDSE simulation with laser
conditions of $\gamma=0.28$ which is in the deep tunneling regime.
It can be seen that below $p_{z}\sim 2 a.u.$, there are several
ring-like interference structures, which means that the direct and
indirect electron trajectories play a dominant role on the total
momentum distribution spectrum, and the interplay between the
intracycle and intercycle interferences induces these ring-like
interference patterns. Such ring-like interference pattern will
blur the fork-like holographic interference structure which can
only be clearly visible at larger $p_{z}$.

\section{IV. CONCLUSION}
In conclusion, we have theoretically investigated 2D photoelectron
momentum distributions in different Keldysh parameter regimes. We
found in deep tunneling regime, profound ring-like interference
pattern can be observed. We have identified that the ring-like
interference pattern is induced by the interplay between the
intra- and inter-cycle interference of electron trajectories. The
center of the ring-like interference pattern lies in where
separations between adjacent temporal double-slit interference
fringes are nearly the same as those of the ATI fringes, which
records electron dynamic information on attosecond sub-cycle
resolution. The appearance of ring-like interference pattern
implies that the rescattered electrons play a negligible role. The holographic
interference pattern can only be visible at larger $p_{z}$, where
the contribution of the rescattered electrons can be discerned. In
the nonadiabatic tunneling regime, the effect of Coulomb potential
will increase, so the contribution of the rescattered electron
trajectories will increase, and the holographic interference
pattern can be clearly distinguished. Our results indicate that
the deep tunneling is not a appropriate condition for observation
of the holographic interference which is more clearly visible
near the nonadiabatic regime.

\section{Acknowledgment}

We benefited from discussions with X. Liu and X. Xie. The work was
supported by the National Basic Research Program of China Grant
(No. 2013CB922201), the NNSF of China (Grant Nos. 11374202,
11274220, 11274050, 11334009 and 11425414), Guangdong NSF (Grant
No. 2014A030311019), and the Open Fund of the State Key Laboratory
of High Field Laser Physics (SIOM). W. Y. is supported by the
``YangFan" Talent Project of Guangdong Province.

\end{document}